\documentclass[twocolumn,aps,pra]{revtex4-1}

\newcommand{\un}[2]{#1\,\mathrm{#2}}

\usepackage{graphicx} 
\usepackage{bm}
\usepackage{natbib}
\usepackage{amsmath}

\newcommand{\eref}[1]{Eq.~(\ref{#1})}

\newcommand{\rtw}{\longrightarrow}

\begin{document}
\preprint{APS/123-QED}

\title{Nuclear charge radii of $^{229}$Th from isotope and isomer shifts}

\author{M. S. Safronova$^{1,2}$, S. G. Porsev$^{1,3}$, M. G. Kozlov$^{3,4}$, J. Thielking$^{5}$, M. V. Okhapkin$^{5}$, P. G\l{}owacki$^{6}$, D. M. Meier$^{5}$, E. Peik$^{5}$}

\affiliation{$^{1}$Department of Physics and Astronomy, University of Delaware, Newark, Delaware 19716, USA\\
$^{2}$Joint Quantum Institute, National Institute of Standards and Technology and the University of Maryland, College Park, Maryland, 20742, USA\\
$^{3}$Petersburg Nuclear Physics Institute of NRC ``Kurchatov Institute'', Gatchina 188300, Russia\\
$^{4}$St. Petersburg Electrotechnical University LETI, St. Petersburg 197376, Russia\\
$^{5}$Physikalisch-Technische Bundesanstalt, 38116 Braunschweig, Germany\\
$^{6}$Pozna\'{n} University of Technology, Pozna\'{n}, Poland}

\begin{abstract}
The isotope $^{229}$Th is unique in that it possesses an isomeric state of only a few eV above the ground state, suitable for nuclear laser excitation. An optical clock based on this transition is expected to be a very sensitive probe for variations of fundamental constants, but the nuclear properties of both states have to be determined precisely to derive the actual sensitivity. We carry out isotope shift calculations in Th$^+$ and Th$^{2+}$ including the specific mass shift, using a combination of configuration interaction and all-order linearized coupled-cluster methods and estimate the uncertainty of this approach. We perform experimental measurements of the hyperfine structure of Th$^{2+}$ and isotopic shift between $^{229}$Th$^{2+}$ and $^{232}$Th$^{2+}$ to extract the difference in root-mean-square radii as $\delta \langle r^{2} \rangle^{232,229}=0.299(15)$ fm$^2$. Using the recently measured values of the isomer shift of lines of $^{229m}$Th, we derive the value for the mean-square radius change between $^{229}$Th and its low lying isomer to be $^{229\textrm{m}}$Th $\delta \langle r^2 \rangle^{229\textrm{m},229} = 0.0105(13)\,{\rm fm}^2$.
\end{abstract}

\maketitle

The nuclear structure of $^{229}$Th is of special interest because of the near-degeneracy of its ground state with an isomer whose energy has been evaluated from differences of $\gamma$ transition energies to be 7.8(5) eV \cite{beck:2009}.
As a result, $^{229}$Th is the only known nucleus with a transition in a laser-accessible region.
 A possibility to drive this nuclear transition coherently with a narrowband laser will open a new regime of precision nuclear spectroscopy, with the application to provide the reference for an optical nuclear clock \cite{peik:2003}.
 In addition to the metrology applications, the development of the nuclear clock is of particular
 interest to searches for physics beyond the standard model of elementary particles due to the potentially
 very high, $|K|=10^4-10^5$,  sensitivity to the variation of the fine-structure constant $\alpha$ \cite{Fla06}. For comparison, the largest
 enhancement among the currently operating atomic clocks is $|K|=6$ \cite{HunLipTam14}. Moreover, the nuclear clock would be sensitive not only to the
 variation of $\alpha$, but also to the variation of ratio of the
 quark masses $m_q$ to the quantum chromodynamics (QCD) scale $\Lambda_{\rm{QCD}}$ \cite{Fla06}, which none of the atomic optical clocks are sensitive to.
 This subject became of even larger interest recently, when the variation of the fundamental constants was directly linked to the
dark matter searches \cite{ArvHuaTil15,DerPos14}. However, the large enhancement factor $K$ for the nuclear clock remains a subject of an open controversy \cite{hayes:2008,litvinova:2009}, which can  be resolved via the determination of the Th nuclear parameters \cite{berengut:2009}, including the change in the
root-mean-square  (rms) charge radius between the isomer and the Th nuclear ground state, which is the goal of the present work.

  General features of the nuclear structure of $^{229}$Th are expected to be similar to other nuclei in the actinide region. They are characterized by a combination of collective rotations and vibrations of the quadrupole-octupole deformed core with the single-particle motion of an unpaired neutron \cite{minkov:2017,bilous:2018,butler:1996}. The ground state rms charge radii of thorium isotopes from 227 to 230 and 232 have been inferred from measured isotope shifts of Th$^+$ \cite{KalRinBek89}.

The direct observation of the  isomeric $^{229\rm{m}}$Th decay was reported by \citet{WenSeiLaa16} and
 the
internal conversion decay half-life of neutral $^{229\rm{m}}$Th  was measured by \citet{seiferle:2017}.
Information on the isomer $^{229\textrm{m}}$Th became available recently from an experiment with trapped $^{229}$Th recoil ions from $\alpha$-decay of $^{233}$U \cite{Thielking:2017}, indicating a small increase of the rms charge radius of the isomer by about 0.001~fm. Theoretical predictions on this number had been disputed in the context of estimating the sensitivity $K$ of the $^{229}$Th nuclear transition frequency to variation of the fine-structure constant. This sensitivity is determined by the change in Coulomb energy $\Delta E_C$  between the ground state and the isomer \cite{hayes:2008,berengut:2009,litvinova:2009}. Predictions vary between $\Delta E_C\approx 0$, expected for negligible coupling between the unpaired neutron and the proton core, and a value on the order of 1 MeV, being required as a Coulomb contribution to compensate similar changes of opposite sign in the contributions from the strong interaction in order to arrive at the small total transition energy of 7.8 eV. Solving this problem based on experimental data on the isomeric charge radius change requires therefore a precise determination of the nuclear radii and moments \cite{berengut:2009}.

 In order to improve the knowledge on the rms radii difference between isotopes of heavy systems, both reliable \textit{ab initio} calculations and precise isotope shift measurements are required. To achieve the required precision, we have carried out the first calculation of the specific mass shift with the high-precision method that combines configuration interaction and the all-order linearized coupled-cluster method.\\

\paragraph*{Theoretical calculation.} We carry out isotope shift (IS) calculations in Th$^+$ and Th$^{2+}$, as well as experimental measurements in Th$^{2+}$ to extract the difference in the rms radii of the $^{229}$Th and $^{232}$Th. These calculations are particularly challenging since Th ions are heavy  systems with mixed electronic configurations containing $5f$ electrons that are very strongly correlated with core electrons. Therefore, both core-valence and valence-valence correlations have to be accounted for with a high precision. To achieve this, we use the hybrid approach that combines configuration interaction (CI) and the all-order linearized coupled-cluster methods [CI+all-order] \cite{SafKozJoh09,OkhMeiPei15}. The isotope shift separates into the mass shift and the field shift (FS). The mass shift is further separated into the normal mass shift (NMS), simply calculated by the scaling of the experimental transition energy and the specific mass shift (SMS), which is very hard to calculate accurately \cite{DzuJohSaf05}.

 The total change in the atomic frequency is given by
\begin{equation}
\Delta \nu^{A^{\prime}A} = (K_{\rm{NMS}}+K_{\rm{SMS}}) \left( \frac{1}{A'} - \frac{1}{A} \right)
+ K^{\prime}_{\rm{FS}}\, \eta^{A'A},
\label{eq11}
\end{equation}
where $\eta^{A'A}$ can be defined (see, e.g.,~\cite{AufHeiSte87}) as
\begin{multline}
\eta^{A'A} \equiv \sum_{k\geq 1} c_k\, \delta \langle r^{2k} \rangle^{A'A}\\
= \delta \langle r^{2} \rangle^{A'A}   \left( c_1 + \sum_{k \geq 2} c_k\, \frac{\delta \langle r^{2k} \rangle^{A'A}}{\delta \langle r^{2} \rangle^{A'A}} \right).
\label{eta}
\end{multline}
Here
\begin{eqnarray}
\delta \langle r^{2k} \rangle^{A'A} &=& \langle r^{2k} \rangle^{A'} - \langle r^{2k} \rangle^{A}
\end{eqnarray}
and $c_k$ are the nuclear parameters.

Assuming that the nucleus can be modeled as a homogeneously charged ball of the radius $R$, it is easy to show that in the linear approximation over $(\delta R)^{A'A} \equiv R^{A'} - R^A$,
\begin{eqnarray}
\delta \langle r^{2k} \rangle^{A'A} = a_k R^{2k-1} (\delta R)^{A'A},
\label{r2k}
\end{eqnarray}
where $a_k$ are the numerical coefficients and $R  \equiv R^{A'} \approx R^A$. As follows from \eref{r2k},
the ratios $\delta \langle r^{2k} \rangle^{A'A}/\delta \langle r^{2} \rangle^{A'A}$ do not depend on $(\delta R)^{A'A}$ in this approximation. Then, introducing the coefficient $K_{\rm{FS}}$, defined as
\begin{equation}
K_{\rm{FS}} = K'_{\rm{FS}} \left( c_1 + \sum_{k=2} c_k\, \frac{\delta \langle r^{2k} \rangle^{A'A}}{\delta \langle r^{2} \rangle^{A'A}} \right),
\end{equation}
we can rewrite \eref{eq11} as follows
\begin{multline}
\Delta \nu^{A^{\prime}A} = (K_{\rm{NMS}}+K_{\rm{SMS}}) \left( \frac{1}{A'} - \frac{1}{A} \right) \\
+ K_{\rm{FS}}\, \delta \langle r^{2} \rangle^{A'A} .
\label{Del_nu}
\end{multline}

Earlier calculations \cite{KalRinBek89,OkhMeiPei15} assumed negligible SMS correction for Th, however we find that it cannot be omitted in a precision calculation. A lowest-order estimate of the SMS correction to the relevant one-electron orbitals in the potential of the Th$^{4+}$ ionic core indicated that SMS can be a few percent of the total IS, requiring further calculations. The NMS correction is only a few MHz for the transitions of interest to this work and is negligible at the present level of accuracy.

The field shift operator, $H_\mathrm{FS}$, modifies the Coulomb potential inside the nucleus. We use the ``finite field'' method, which means that perturbation is added to the initial Hamiltonian with the arbitrary coefficient $\lambda$: $H\rtw H_\lambda= H + \lambda H_\mathrm{FS}$. The coefficient $\lambda$ has to be sufficiently large  to make the effect of the field shift significantly larger than the numerical uncertainty of the calculations but small enough to keep the change in the energy linear with $\lambda$. We find eigenvalues $E$ by direct diagonalization of $H_\lambda$ and then find field shift coefficient $K_{\mathrm{FS}}$ as \cite{KK07,KozKor05}:
\begin{equation}
 K_\mathrm{FS} = \frac{5}{6R^2} \frac{\partial E}{\partial\lambda} .
\label{der}
\end{equation}
It was verified that using the Fermi distribution does not change the results well within the uncertainties of the calculations~\cite{KalBehGor17}.

\begin{figure}[ht]
\centering
\includegraphics[width=0.40\textwidth]{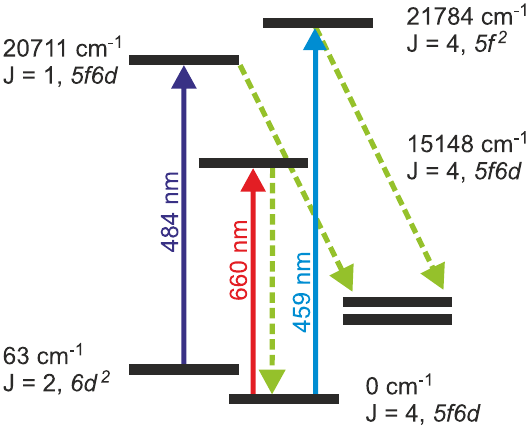}
\caption{\label{fig:isotope} Partial level scheme of Th$^{2+}$, showing the transitions relevant to our investigations. The levels are labeled by their energy in cm$^{-1}$, electronic angular momentum $J$ and the dominant configuration of the valence electrons. Laser excitation is shown with solid arrows. The detection is provided via fluorescence decay, indicated by dashed lines.}
\end{figure}

\begin{figure}[ht]
\centering
\includegraphics[width=0.49\textwidth]{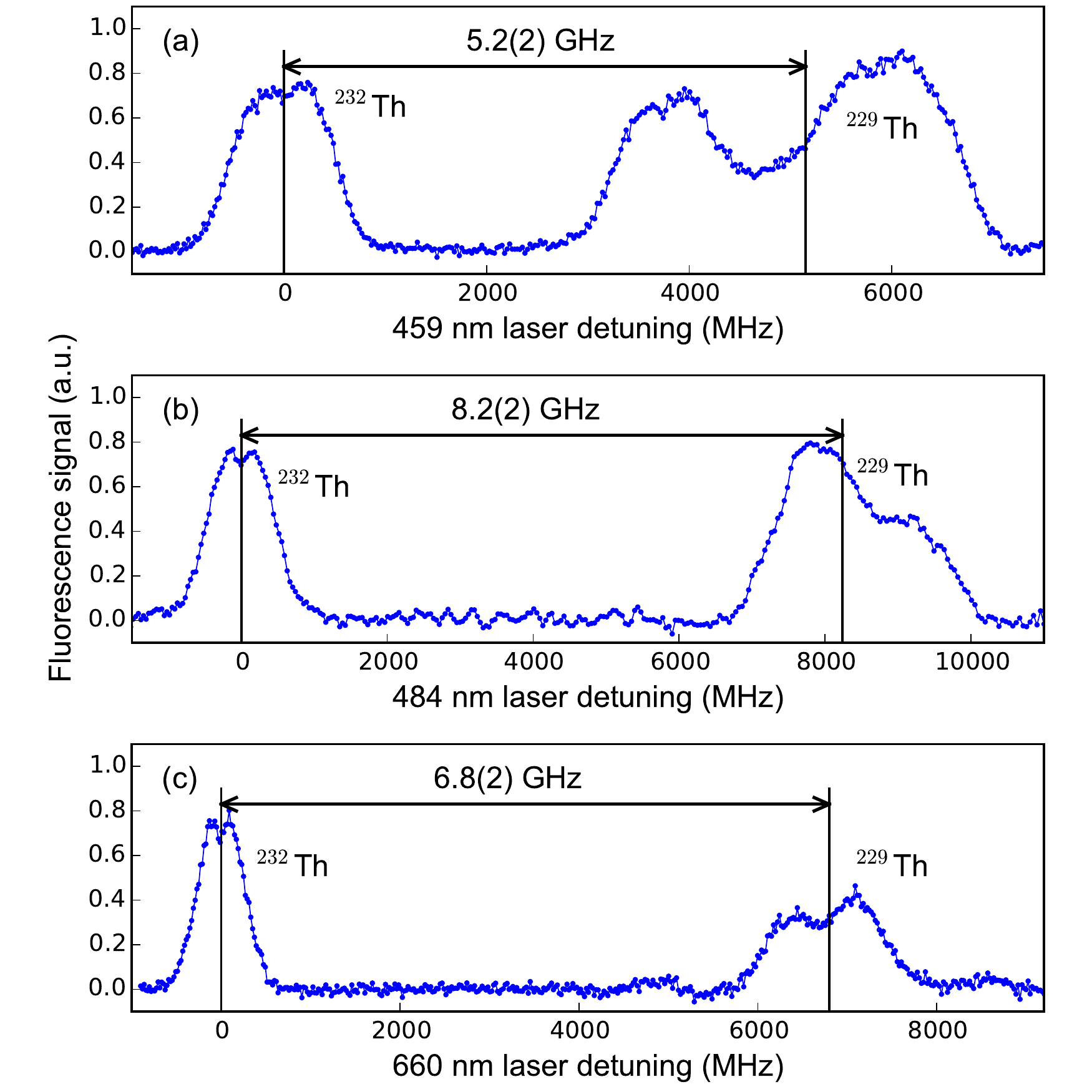}
\caption{\label{fig:spectra} Fluorescence signals of $^{232}$Th$^{2+}$ and $^{229}$Th$^{2+}$ obtained with saturated absorption spectroscopy. The Doppler-free resonances are visible in the $^{232}$Th$^{2+}$ resonances, but most of them are not resolved for $^{229}$Th$^{2+}$ due to the limited signal-to-noise ratio. The centers of the hyperfine structures of $^{229}$Th$^{2+}$ are therefore calculated using the hyperfine constants presented in ~\cite{Mueller:2018}.  a) Transition $0_{4}\rightarrow 21784_{4}$ at 459 nm, b) transition $63_{2}\rightarrow 20711_{1}$ at 484 nm, c) transition $0_{4}\rightarrow 15148_{4}$ at 660 nm.}
\end{figure}

The conversion factor from atomic units to SI units for the coefficient $K_\mathrm{FS}$  is $1\, \mathrm{a.u.} = 2.3497\cdot 10^{-3}\, {\mathrm{GHz}}/{\mathrm{fm}^2}$. The specific mass shift is calculated by modifying the Hamiltonian with the SMS operator $H\rtw H_\lambda= H + \lambda H_\mathrm{SMS}$. The SMS coefficient is given by the corresponding derivative $K_\mathrm{SMS} = \frac{\partial E}{\partial\lambda}.$ The conversion factor from atomic units to SI units for the coefficient $K_\mathrm{SMS}$  is $1\, \mathrm{a.u.} = 3609.46\, {\mathrm{GHz}}/{\mathrm{amu}}$.\\
\paragraph*{Experimental method.}
In the experiment we use a radio-frequency trap ~\cite{Herrera:2012} loaded with $\approx 10^{6}$ Th$^{+}$ ions by laser ablation from a target containing $^{229}$Th and $^{232}$Th and further three-photon ionization of trapped Th$^+$ to produce $\approx 10^{3}$ Th$^{2+}$ ions~\cite{Thielking:2017}. The ions are cooled close to room temperature by collisions with a high-purity argon buffer gas. The isotopic shift between $^{229}$Th and $^{232}$Th is measured for three transitions in Th$^{2+}$ from the electronic ground state and the low-lying $63_{2}$ state (levels are labeled by their energy in cm$^{-1}$ and electronic angular momentum $J$ as subscript, see Fig.~\ref{fig:isotope}). We excite the transitions from the state $0_{4}$ to the states $15148_{4}$ and $21784_{4}$ with external-cavity diode lasers (ECDL) at $\un{660}{nm}$ and $\un{459}{nm}$ respectively and the transition from the level $63_{2}$ to $20711_{1}$ with an ECDL at $\un{484}{nm}$.
The $63_{2}$ state is populated by about ${30}\%$ of the ions in the trap, which provides suitable excitation signals. The output power of all lasers is in the range of $\un{10}{mW}$. We scan the frequency of the ECDLs  in the range of $\approx\un{10}{GHz}$ to cover the frequency interval between the $^{232}$Th and $^{229}$Th isotopes. The laser beams are retro-reflected to provide saturated absorption spectroscopy. To measure the frequency detuning of the lasers during the scanning, a temperature stabilized confocal cavity is used. The fluorescence detection is provided by a photomultiplier operating in photon counting mode. For the excitations at $\un{459}{nm}$ and $\un{484}{nm}$ sensitive fluorescence detection of these ions is performed using decay channels at other wavelengths, free from background of laser stray light. The fluorescence signal for the excitation at $\un{660}{nm}$ is detected on the same wavelength and therefore is not free from laser background. The three spectra are shown in Fig.~\ref{fig:spectra}. Most of the Doppler-free resonances of the $^{229}$Th$^{2+}$ HFS are not resolved due to insufficient signal-to-noise ratio, limited by the ion storage time. The resolution is therefore limited to the Doppler-broadened absorption linewidth in the range of $\un{750}{MHz}$. To identify the center of each HFS of $^{229}$Th$^{2+}$, we fit the shape of the HFS using the hyperfine constants presented in ~\cite{Mueller:2018} and calculate the frequency for $A=B=0$.
\\
\paragraph*{Results.}
\begin{table*}[t]
\caption{\label{tab1} Field shift and specific mass shifts calculations in Th$^+$ and Th$^{2+}$ transitions and the
extraction of $\delta \langle r^{2} \rangle^{232,229}$
Th$^+$ experimental value is from \cite{KalRinBek89}. Experimental energies from \cite{expt} are given for reference. }
\begin{ruledtabular}
\begin{tabular}{lllccccccccc}
\multicolumn{1}{c}{Ion} &
\multicolumn{1}{c}{Energy} &
\multicolumn{1}{c}{Electronic} &
\multicolumn{4}{c}{Field shift} &
\multicolumn{3}{c}{Specific mass shift} &
\multicolumn{1}{c}{IS Expt.} &
\multicolumn{1}{c}{$\delta \langle r^{2} \rangle^{232,229}$} \\
\multicolumn{1}{c}{} &
\multicolumn{1}{c}{Ref.~\cite{expt}.} &
\multicolumn{1}{c}{configuration} &
\multicolumn{1}{c}{$\partial E/\partial \lambda$} &
\multicolumn{1}{c}{$\partial E/\partial \lambda$} &
\multicolumn{1}{c}{} &
\multicolumn{1}{c}{$K_{\rm{FS}}$} &
\multicolumn{1}{c}{$\partial E/\partial \lambda$} &
\multicolumn{1}{c}{$K_{\rm{SMS}}$} &
\multicolumn{1}{c}{$\delta \nu_{\rm{SMS}}$} &
\multicolumn{2}{c}{} \\
\multicolumn{3}{c}{} &
\multicolumn{1}{c}{CI+MBPT} &
\multicolumn{1}{c}{CI+All} &
\multicolumn{1}{c}{Diff.} &
\multicolumn{1}{c}{CI+All} &
\multicolumn{1}{c}{CI+All} &
\multicolumn{1}{c}{CI+All} &
\multicolumn{1}{c}{CI+All} &
\multicolumn{2}{c}{}  \\
\multicolumn{1}{c}{} &
\multicolumn{1}{c}{cm$^{-1}$} &
\multicolumn{1}{c}{} &
\multicolumn{1}{c}{a.u.} &
\multicolumn{1}{c}{a.u.} &
\multicolumn{1}{c}{\%} &
\multicolumn{1}{c}{GHz/fm$^2$} &
\multicolumn{1}{c}{a.u.} &
\multicolumn{1}{c}{GHz/amu} &
\multicolumn{1}{c}{GHz} &
\multicolumn{1}{c}{GHz}&
\multicolumn{1}{c}{fm$^2$} \\ [0.3pc]
\hline
Th$^{2+}$&    63 & $6d^2$~$^3\!F_2$&  -0.000421  & -0.000372 & 13\% & -36.6 & -1.79 &  -6440 & 0.364  &        & \\
          & 20711 & $5f6d$~$^1\!P_1$&  -0.000667  & -0.000647 &  3\% & -63.6 & -3.30 & -11900 & 0.672  &        & \\
          & 29300 & $5f^2$~$J=0$    &  -0.000805  & -0.000854 &  6\% & -84.1 & -4.19 & -15100 & 0.853  &        & \\ [0.1pc]
 Transitions         &       & $6d^2 - 5f6d$   &  0.000247   &  0.000274 & 10\% &  27.0 & 1.52 &  5480 & -0.309  & 8.2(2) & 0.315(32)\\
          &       & $5f6d - 5f^2$   &  0.000138   &  0.000208 & 34\% &  20.4 & 0.88 &  3180 & -0.180   & 6.2(3) & 0.312(42)\\[0.5pc]

 Th$^{2+}$ &0     &$5f6d$~$^3\!H_4$&  -0.0007129 &	-0.0006981&	2\%	 & -68.7& -3.34	&-12000 &	0.680	&	&\\
           &15149 &$5f^2$~$^3\!H_4$&  	         &  -0.0009096&		 & -89.5& -4.47	&-16100 &	0.912	&	&\\
           &21784 &$5f^2$~$^3\!F_4$&  -0.0008133 &	-0.0008683&	6\%	 & -85.5& -4.31	&-15550 &	0.878	&	&\\ [0.1pc]
Transitions           &	  &$ 5f6d - 5f^2$~$^3\!H_4$ &	         &	 0.0002115&	  	 & 20.8	&  1.14	&4100	&  -0.232	&6.8(2)&	0.335(43)\\
           &	  &$ 5f6d - 5f^2$~$^3\!F_4$  &  0.0001004	 &   0.0001703&	41\% & 16.8	&  0.97	&3510	&  -0.198	&5.2(2)&	0.332(54)\\ [0.5pc]
							
Th$^+$    &     0 & $6d7s^2$~$^2\!D_{3/2}$& 0.000572  &  0.000555 &  3\% &54.6 & -2.24 &  -8090 & 0.457  &        &        \\
          & 17122 & $5f6d^2$~$J=3/2$      & -0.000360  & -0.000322 & 12\% & -31.6 & -3.69 & -13300 & 0.751  &        &         \\ [0.1pc]
Transition          &       & $6d7s^2 -5f6d^2$     &  0.000932   &  0.000876 &  6\% & 86.2 & 1.45 &  5240 & -0.296  &25.01(9)& 0.294(17)\\
 Final    &       &                       &            &           &      &      &      &       &         &        & 0.299(15)\\
\end{tabular}
\end{ruledtabular}
\end{table*}
The results for the FS and SMS ${\partial E}/{\partial\lambda}$ derivatives and coefficients are summarized in Table~\ref{tab1}. We use two different methods for the calculations of the field shift: combination of the CI  with many-body perturbation theory (CI+MBPT) \cite{DzuFlaKoz96} and more accurate CI+all-order method \cite{SafKozJoh09}. These approaches allow one to incorporate core excitations in the CI method by constructing an effective Hamiltonian $H_{\mathrm{eff}}$ using either second-order MBPT or linearized coupled-cluster methods, respectively. The CI+all-order method includes third- and higher-order corrections to the effective Hamiltonian. Using two methods allows to establish the effect of the higher orders and to estimate the accuracy of the final results. The corresponding results for the derivatives are given in the columns labeled CI+MBPT and CI+All. The difference between these results gives an  estimate of the uncertainty of our calculation, listed in the ``Diff.'' column in \%. For the $5f^2$~$^3\!H_4$ state, the CI+MBPT approximation gives incorrect level mixing with the $J=4$ even states leading to a poor result for the IS.

We note that we use a much larger set of the configurations in the CI calculation in comparison with \cite{OkhMeiPei15}. The number of configurations was increased to ensure negligible numerical uncertainty in the CI calculation.

The SMS in GHz is listed in the  $\delta \nu_{\rm{SMS}}$ column. The SMS is 3-4\% of the total IS in two Th$^{2+}$ transitions of interest but only 1\%  for the Th$^{+}$ transition listed in Table~\ref{tab1}. Taking into account that the SMS is small, we calculate it only in the CI+all-order approximation. We roughly estimate its uncertainty as the difference of the one-electron SMS for averaged $6d-5f$ difference, (-0.366 GHz), and final CI+all-order values. Thus, the uncertainties for the $5f6d - 6d^2$ Th$^{2+}$ and $5f6d^2 - 6d7s^2$ Th$^+$ transitions are $\sim$ 20\%.

The value of $\delta \langle r^{2} \rangle^{232,229}$ is extracted by combining experimental and theoretical values according to Eq.~(\ref{Del_nu}). The theoretical  result for the $6d7s^2 - 5f6d^2$ transition in Th$^+$ ion is more accurate, the uncertainty being about 6\%, because the FS  shifts the levels  in the opposite directions. As a result, there is no cancellation between upper and lower levels FS, leading to higher accuracy  or this case. We estimate the uncertainty in the  $6d^2$~$^3\!F_2 - 5f6d~^1\!P_1$ IS in Th$^{2+}$ to be 10\% based on the difference of the CI+MBPT and CI+all-order results. In the other three cases, this procedure is expected to significantly overestimate the uncertainty due to poor approximation given by the CI+MBPT.  Therefore, we use the absolute uncertainly in the FS constant, 2.7 GHz/fm$^2$, for the  $6d^2$~$^3\!F_2 - 5f6d~^1\!P_1$ transition as the uncertainty for the other three Th$^{2+}$ $6d-5f$ transitions.

The final value is the weighted average of the two results obtained for the $5f6d - 6d^2$ Th$^{2+}$ and $5f6d^2 - 6d7s^2$ Th$^+$ transitions. The weights are calculated as the inverse of the squares of the uncertainties in the $\delta \langle r^{2} \rangle^{232,229}$ values. When calculating the weighted average, we do not include the results for the other transitions in Th$^{2+}$ since we cannot reliably estimate their uncertainties.  However, the rms results extracted from all transitions are consistent well within the estimated uncertainties.

The final result is $\delta \langle r^{2} \rangle^{232,229}=0.299(15)$ fm$^2$. This result is in a good agreement with the present value 0.33(5) fm$^2$~\cite{KalRinBek89} but is 3 times more accurate. A compilation of data on nuclear ground state charge radii \cite{AngMar13} gives $\delta \langle r^{2} \rangle^{232,229}=0.334(8)$ fm$^2$, relying on Ref. \cite{KalRinBek89} as reference data for thorium, combined with a fit over isotopic sequences of other elements.

We can now give an updated value for the mean-square radius change between $^{229}$Th and its low lying isomer $^{229\textrm{m}}$Th, since it is derived from $\delta \langle r^{2} \rangle^{232,229}$ and the ratio of the isomeric line shift and the isotopic shift between $^{232}$Th and $^{229}$Th. Using the ratio of the isomeric and isotopic shifts given in \cite{Thielking:2017} and our value for $\delta \langle r^{2} \rangle^{232,229}$, we obtain $\delta \langle r^2 \rangle^{229\textrm{m},229} = 0.0105(13)\,{\rm fm}^2$.

The smallness of the difference of the rms charge radii between isomer and ground state makes it challenging to determine the difference in Coulomb energy as  proposed in \cite{berengut:2009} because quadrupole and higher-order deformations lead to significant contributions. We have performed numerical calculations of the Coulomb energy in the liquid drop model for shapes in the range of the $^{229}$Th ground state deformations $\beta_2=0.2$, $\beta_3=0.1$, and $\beta_4=0.1$ \cite{moeller:1995, butler:1996}. The present uncertainty from the spherical contribution $\delta \langle r^2 \rangle^{229\textrm{m},229}$ to the Coulomb energy difference is  about 30 keV. In order to reach a similar uncertainty for the deformed nucleus (regardless of uncertainty from the nuclear model) the differences in the three parameters $\beta_2,~\beta_3$ and $\beta_4$  will have to be determined experimentally with uncertainties in the low $10^{-3}$ range. This emphasizes the interest in precision studies of the  $^{229}$Th$^{3+}$ hyperfine structure, including higher-order contributions beyond the electric quadrupole \cite{beloy:2014}.

\begin{acknowledgments}
We acknowledge financial support from the European Union's Horizon 2020 Research and Innovation Programme under Grant Agreement No. 664732 (nuClock) and from DFG through CRC 1227 (DQ-mat, project B04). This work was supported in part by the Office of Naval
Research, USA, under award number N00014-17-1-2252 and Russian Foundation
for Basic Research under Grant No. 17-02-00216.
\end{acknowledgments}

%

\end{document}